\documentclass[graybox, envcountchap, authoryear, natbib]{svmult}

\usepackage{amsmath}
\usepackage{amssymb}

\usepackage{graphicx}        

\begin{document}

\title*{The Effect of Galactic Feedback on Gas Accretion and Wind Recycling} 
\author{Freeke van de Voort}
\institute{Freeke van de Voort \at Department of Astronomy and Theoretical Astrophysics Center, University of California, Berkeley, CA 94720-3411, USA \\ Academia Sinica Institute of Astronomy and Astrophysics, P.O. Box 23-141, Taipei 10617, Taiwan \\ Heidelberg Institute for Theoretical Studies, Schloss-Wolfsbrunnenweg 35, 69118, Heidelberg, Germany \\ Astronomy Department, Yale University, PO Box 208101, New Haven, CT 06520-8101, USA \\ \email{freeke.vandevoort@h-its.org}}

\maketitle

\abstract{In the absence of galactic winds, the rate at which gas accretes onto galaxies is determined by the gravitational potential and by radiative cooling. However, outflows driven by supernovae and active galactic nuclei not only eject gas from galaxies, but also prevent gas from accreting in the first place. Furthermore, gas previously ejected from a galaxy can re-accrete onto (the same or a different) galaxy. Because this gas has a high metallicity, its cooling rate is relatively high, which will increase its chances to re-accrete. This complex interplay between gas inflows and outflows is discussed in this chapter. Wind recycling is found to be an important process that fuels galaxies at late times and the recycled gas has different properties than gas accreting for the first time. Quantitative conclusions, however, vary between studies, because the amount of wind recycling is dependent on the details of the feedback model. We discuss these differences, known caveats, and ways to make progress in understanding how galaxies are fed at low redshift.}

\section{Introduction}

Structure formation in the Universe is governed by gravity. Observations of the cosmic microwave background radiation allow us to measure small fluctuations in temperature, which correspond to small fluctuations in density. These fluctuations grow through gravitational collapse, forming bound structures like galaxies and stars. In the standard cosmological constant cold dark matter ($\Lambda$CDM) model, mass assembles hierarchically, with the smallest structures forming first. As the matter in the Universe collapses, it forms a network of sheets and filaments, the so-called ‘cosmic web’. While the collapse of dark matter halts as it reaches virial equilibrium in near-spherical halos, baryons can radiate away their binding energy, allowing them to collapse further and fragment into smaller structures. Galaxies form in the densest regions and the most massive galaxies form where filaments intersect. The evolution of galaxies is critically linked to the cosmic web and gas flows in their halos.

The formation of galaxies and, specifically, the accretion of gas onto galaxies are complex problems. There is a variety of evidence showing that galaxies do not sit passively in the densest regions of the cosmic web, but instead have a profound impact on their surroundings. The intergalactic medium (IGM) is enriched with metals, as shown by observations of metal absorption lines in quasar spectra \citep[e.g.][]{Peroux2005, Turner2016}. The baryon fraction in galaxy groups, as determined from X-ray observations of the hot gas around these galaxies, is much lower than the cosmic mean baryon fraction \citep[e.g.][]{Sun2009, Gonzalez2013}. The observed cosmic star formation rate (SFR) density is much below the one predicted by cosmological, hydrodynamical simulations without galactic winds \citep[e.g.][]{Schaye2010}. 

This indicates that winds emanating from galaxies remove gas from the interstellar medium (ISM) and change the properties of the circumgalactic medium (CGM) and IGM by polluting it with metals and potentially heating it. However, gas inflow and outflow cannot be viewed separately. They interact with each other gravitationally and, more importantly, hydrodynamically. Outflows will plow into static or inflowing gas and sweep up extra material on their way out. This inevitably slows down the outflow until it either joins the warm-hot intergalactic medium \citep{Dave2001} or forms cold gas clouds through efficient cooling. It can then possibly reverse its trajectory and re-accrete onto the galaxy. This is known as wind recycling and sometimes referred to as a halo fountain or intergalactic fountain \citep{Oppenheimer2008, Oppenheimer2010, Keres2009b}. There is no clear distinction between wind recycling and galactic fountains, but in the case of wind recycling the gas reaches larger distances, out in the halo, before rejoining the galaxy. Another form of gas recycling happens through stellar mass loss, where gas that was previously locked up in stars becomes part of a stellar wind, rejoins the ISM, and can collapse again to form stars \citep[see e.g.][and references therein]{Segers2016}. This chapter, however, is concerned with wind recycling between galaxies and their halos. 

\begin{figure}[!ht]
\sidecaption 
\includegraphics[scale=.35]{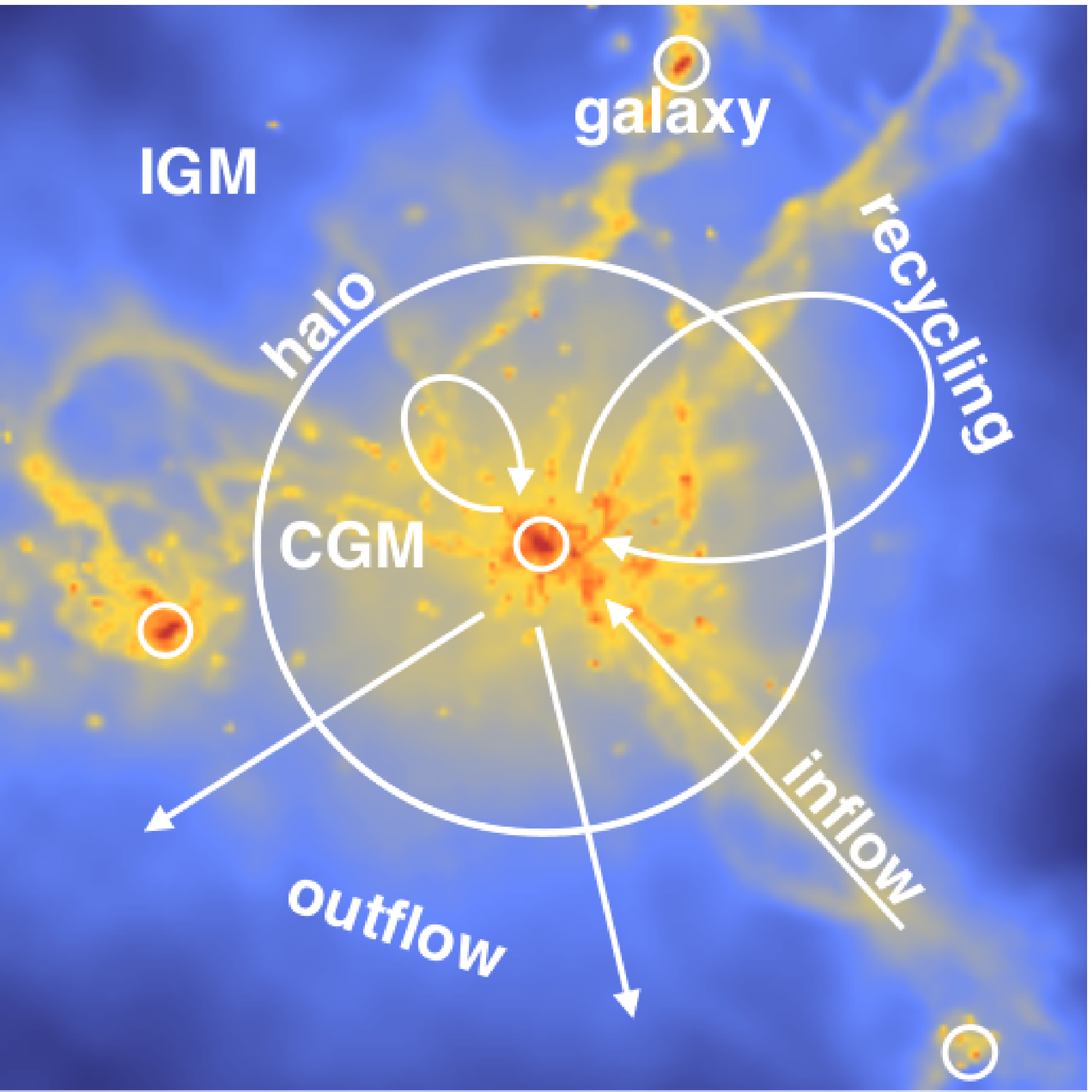}
\caption{\label{fig:halo} Gas density in a (450~kpc)$^3$ volume centered on a $10^{12}$~M$_\odot$ halo at $z=2$. The large white circle indicates the virial radius and the small white circles show the locations of several galaxies. Gas inflow happens primarily via dense filaments connected to the cosmic web. Outflows escape along the path of least resistance in underdense regions. Wind recycling can take place within the halo or the gas can escape the halo and re-accrete on a longer time-scale. Adapted from \citet{VoortSchaye2012}.}
\end{figure}
Figure~\ref{fig:halo} shows the gas density in a (450~kpc)$^3$ volume in and around a $10^{12}$~M$_\odot$ halo at $z=2$. The halo shown is being fed primarily by dense, clumpy filaments and as well as by external galaxies embedded in the filaments. Gas outside the halos is, in general, moving towards the halo and relatively unenriched. Within the virial radius, however, most of the underdense regions are filled with diffuse, outflowing gas. The inflowing gas is generally associated with the dense, cold streams, which can penetrate the halo and feed the central disc, although the hot, diffuse gas can also cool and accrete onto the galaxy \citep[e.g.][]{Keres2005,Voort2011a}. At lower redshift, filaments are less pronounced and cooling from the hot halo is fractionally more important. The virial radius only approximately marks the boundary between the unaffected IGM and the enriched and heated CGM and many studies find effects of galaxy formation out to $2R_\mathrm{vir}$ or beyond \citep[e.g.][]{Wetzel2012, Rasmussen2012, Bahe2013}. 

Satellites contribute directly to the growth of the central galaxy through merging. Gas previously processed in and ejected or stripped from companion galaxies can also accrete onto the main galaxy. It is a matter of definition whether or not to consider this gas wind recycling. One argument in favor of including it is that the gas has had its properties altered through feedback. Not including it may be a better choice when interested in the return of gas to the same galaxy in order to quantify the efficiency of galactic winds. 

This chapter discusses our current understanding of how gas accretes and re-accretes onto galaxies and how feedback affects the accretion rate and changes the properties of the galaxy and its CGM. Section~\ref{sec:virial} discusses some theoretical background of galaxy and halo formation that is necessary to understand gas accretion. We discuss three methods in Section~\ref{sec:methods} that are used to study galaxy formation, semi-analytic models, equilibrium models, and hydrodynamical simulations. We focus on hydrodynamical simulations in Section~\ref{sec:results} and describe how feedback affects the accretion rate onto galaxies and properties of the accreting gas and how important wind recycling is. Finally, we compare and discuss the results from the literature and conclude in Section~\ref{sec:concl}.

\section{Virial relations} \label{sec:virial}

The following discusses a highly idealized, but nonetheless useful case of spherical collapse in a flat universe, where the radiation density parameter $\Omega_\gamma$, the matter density parameter $\Omega_\mathrm{m}$, and the cosmological constant or vacuum density parameter $\Omega_\mathrm{\Lambda}$ add up to one, $\Omega_\gamma + \Omega_\mathrm{m} + \Omega_\mathrm{\Lambda}=\Omega=1$.
\begin{equation}
\Omega \equiv \dfrac{\langle\rho\rangle}{\rho_\mathrm{crit}}=\dfrac{8\pi G\langle\rho\rangle}{3H^2},
\end{equation}
with $\rho_\mathrm{crit}$ the redshift-dependent density at which the universe is flat, or `critically bound'. $G$ is the gravitational constant, $\langle\rho\rangle$ the mean density of the universe, and $H$ the Hubble constant. The mean cosmic matter density is
\begin{equation}
\langle\rho_m\rangle =  \Omega_\mathrm{m}\rho_\mathrm{crit} = \Omega_\mathrm{m}(1+z)^3\rho_\mathrm{crit,0} =\Omega_\mathrm{m}(1+z)^3\dfrac{3H_0^2}{8\pi G},
\end{equation}
where the subscript 0 is used to indicate the present-day value.
The density at which objects collapse or virialize can be calculated analytically for an Einstein-de Sitter universe \citep{Padmanabhan2002}, but it holds for more general cosmologies \citep{Bryan1998}.
\begin{equation}
\rho_\mathrm{coll}\approx18\pi^2\langle\rho_m\rangle
\end{equation}
Assuming spherical collapse, the mass of this halo then becomes
\begin{equation}
M_\mathrm{halo}=\dfrac{4\pi}{3}\rho_\mathrm{coll}R_\mathrm{vir}^3.
\end{equation}
This leads to an expression for the virial radius
\begin{equation} \label{eqn:radius}
R_\mathrm{vir}\approx\left(\dfrac{2GM_\mathrm{halo}}{H_0^2\Omega_\mathrm{m}18\pi^2}\right)^{1/3}\dfrac{1}{1+z}
\end{equation}
or 
\begin{equation}
R_\mathrm{vir}\approx3.4\times10^2\ \mathrm{kpc}\ \left(\dfrac{M_\mathrm{halo}}{10^{12}\ \mathrm{M}_\odot}\right)^{1/3}\dfrac{1}{1+z}.
\end{equation}
$R_\mathrm{vir}$ approximately marks the location of an accretion shock, within which the shocked gas is in virial equilibrium with the dark matter.

The virial theorem states that the total kinetic energy, $K$, of a system is equal to minus one half of the average total potential energy, $U$.
\begin{equation} \label{eqn:virial}
K=-\frac{1}{2}U
\end{equation}
Taking $K=\frac{1}{2}mv_\mathrm{c}^2$ and $U=-\dfrac{GM_\mathrm{halo}m}{R_\mathrm{vir}}$, with $m$ the mass of a test particle, the circular velocity, $v_\mathrm{c}$, of the system is
\begin{equation}
v_\mathrm{c}^2 = \dfrac{GM_\mathrm{halo}}{R_\mathrm{vir}}.
\end{equation}
In a monatomic ideal gas, the thermal energy of the particles is equal to their kinetic energy
\begin{equation} \label{eqn:thermal}
\frac{3}{2}k_\mathrm{B}T=\frac{1}{2}\mu m_\mathrm{H}v_\mathrm{c}^2,
\end{equation}
where $k_\mathrm{B}$ is Boltzmann's constant and $m_\mathrm{H}$ the mass of a proton. For a fully ionized plasma with primordial composition, the mean molecular weight $\mu\approx0.59$. Adding heavier elements does not change this quantity significantly. Combining equations \ref{eqn:virial} and \ref{eqn:thermal} gives an expression for the virial temperature of a monatomic ideal gas
\begin{equation}
T_\mathrm{vir} = \dfrac{\mu m_\mathrm{H}GM_\mathrm{halo}}{3k_\mathrm{B} R_\mathrm{vir}}.
\end{equation}
Filling in the expression for the virial radius (equation \ref{eqn:radius}) one gets
\begin{equation} \label{eqn:virialtemperature}
T_\mathrm{vir}=\left(\dfrac{G^2H_0^2\Omega_\mathrm{m}18\pi^2}{54}\right)^{1/3}\dfrac{\mu m_\mathrm{H}}{k_\mathrm{B}}M_\mathrm{halo}^{2/3}(1+z)
\end{equation}
or
\begin{equation}
T_\mathrm{vir}\approx 3.0\cdot10^5\ \mathrm{K}\ \left(\dfrac{\mu}{0.59}\right)\left(\dfrac{M_\mathrm{halo}}{10^{12}\ \mathrm{M}_\odot}\right)^{2/3}\left(1+z\right).
\end{equation}

There is an uncertainty of a factor of a few because of the assumption we made for the average total potential energy.
The virial temperature of a halo is completely determined by its mass and redshift, for given cosmological parameters. Due to adiabatic compression, the temperature of the hot, quasi-hydrostatic halo gas can increase for $R<R_\mathrm{vir}$, but this is only a minor effect. Feedback from supernovae or supermassive black holes can also heat the gas to above $T_\mathrm{vir}$ \citep[e.g.][]{Voortetal2016, Fielding2016}. 

The gas density, $\rho$, increases by orders of magnitude from $R_\mathrm{vir}$ towards the center of the halo.  Assuming that the density profile is approximately self-similar for different halo masses, it follows that the pressure of the gas also increases roughly as $M_\mathrm{halo}^{2/3}$ at fixed $R/R_\mathrm{vir}$, since $P=k_\mathrm{B}T\rho/\mu m_\mathrm{H}$. At fixed radius $R$, the pressure scales linearly with $M_\mathrm{halo}$. It is therefore expected that galactic winds, if launched at the same velocity, slow down more quickly in more massive halos, partially because of their deeper potential wells, but also because of the much higher ram pressure forces.

\subsection{Cooling time}

Gas falling towards a galaxy gains kinetic energy and in order for it to join the galaxy's ISM, it has to lose this excess kinetic energy. If the gas' velocity is supersonic when it reaches the hydrostatic halo, it will experience a shock and heat to the virial temperature of the halo. According to the simplest picture of spherical collapse, all gas in a dark matter halo is heated to the virial temperature of that halo, reaching a quasi-static equilibrium supported by the pressure of the hot gas. 

Within the so-called cooling radius, the cooling time of the gas will, however, be shorter than the age of the Universe. If the cooling radius lies well inside the halo, which is the case for high-mass halos, a quasi-static, hot atmosphere will indeed form. Accretion onto the galaxy is then regulated by the cooling function. If, on the other hand, this radius is larger than the virial radius, then there will be no hot halo and the gas will not go through an accretion shock at the virial radius \citep{Rees1977, White1978, Birnboim2003, Keres2005}. The accretion rate onto the central galaxy then depends on the infall rate, but not on the cooling rate \citep{White1991}. A clumpy or filamentary structure of the IGM and CGM will change this simplified picture and result in only part of the infalling gas shock heating at $R_\mathrm{vir}$ \citep{Keres2005, Dekel2006} 

Gas with a temperature below about $10^6$~K can cool down rapidly, through line emission. Cooling occurs mostly through particle collisions and the cooling rate therefore depends on the number densities of both the colliding particle species. These number densities depend on the density of the gas. The cooling time depends on the density via
\begin{equation}
t_\mathrm{cool}=\dfrac{u}{\Lambda}\propto\dfrac{\rho}{\rho^2}=\dfrac{1}{\rho}
\end{equation}
where $t_\mathrm{cool}$ is the cooling time, $\Lambda$ the cooling function, and $u$ the thermal energy density, equal to $\dfrac{3}{2}nk_BT$ and therefore proportional to the density of the gas. The cooling time is thus inversely proportional to the density of the cooling gas.

\begin{figure}[!ht]
\center 
\includegraphics[scale=.5]{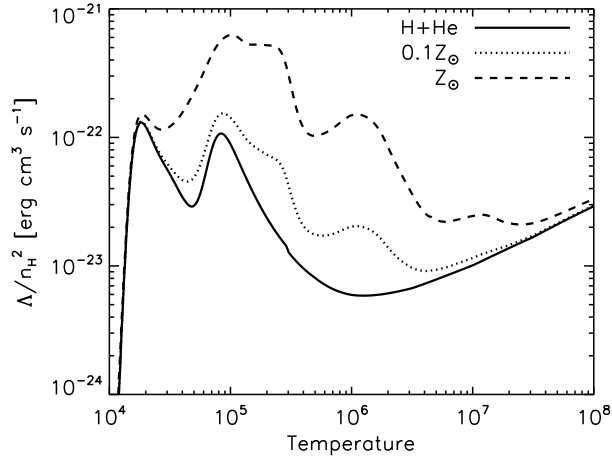}
\caption{\label{fig:cool} Cooling function for optically thin gas with hydrogen number density $n_\mathrm{H}=0.001$~cm$^{-3}$ in photo-ionization equilibrium with the \citet{Haardt2001} UV background. Adapted from \citet{Wiersma2009a}.}
\end{figure}
The dependence of the cooling function on temperature and metallicity is shown in Figure~\ref{fig:cool}. The curves show the cooling function for an optically thin gas with hydrogen number density $n_\mathrm{H}=0.001$~cm$^{-3}$ in photo-ionization equilibrium with the \citet{Haardt2001} UV background. Different curves have different metallicities, primordial (solid), 10 per cent of the solar metallicity (dotted), and solar metallicity ($Z_\odot$; dashed). At $Z_\odot$ cooling through metal-line emission dominates except at $T\approx10^4$~K, where hydrogen lines are most important, and at $T>10^7$~K, where Bremsstrahlung takes over. Cooling through metal lines becomes less important with decreasing metallicity and at $0.1Z_\odot$ changes $\Lambda$ by less than a factor of two at most temperatures. At higher redshift, the density of the CGM is higher and the average cooling time is therefore shorter. Because galactic feedback changes the temperature, density, and metallicity of the CGM, it has a strong effect on the cooling of the halo gas. Because gas that is ejected from a galaxy has a higher metallicity than average, it will cool more efficiently (at fixed temperature and density) and possibly re-accrete onto the galaxy.

\section{Methods to probe the gas cycle} \label{sec:methods}

Observations of groups of galaxies have revealed that their gas and baryon fractions are substantially below the cosmic mean baryon fraction, whereas clusters of galaxies have fractions much closer to the cosmic mean \citep[e.g.][]{Sun2009, Vikhlinin2009, Gonzalez2013}. The reason for this is likely that feedback from galaxies removes a substantial fraction of gas from group-sized halos. Simulations predict that the baryon fraction decreases further towards lower halo mass, where it is easier for stellar feedback to expel gas from the halo \citep[e.g.][]{Chan2015, Christensen2016, Voortetal2016}. The gas and baryon fractions are also lower at lower redshift, after powerful outflows remove a large amount of gas from the halo \citep{Muratov2015, Voortetal2016}. This results in halos of different mass not being self-similar and also directly effects observables such as the X-ray luminosity and the magnitude of the Sunyaev-Zel'dovich effect \citep{Crain2010a, Brun2014, Voortetal2016}.

\subsection{Semi-analytic models}

The technique of modeling galaxy evolution on top of a dark matter-only N-body simulation is known as `semi-analytic'. Semi-analytic models combine dark matter halo merger trees with analytic recipes for baryonic processes. A clear advantage over doing full hydrodynamical simulations is their lower computational cost. Galaxy formation is too complex to construct an ab initio model. Therefore, it is possible to conduct a large parameter study in order to test how observations of, for example, the galaxy stellar mass function can best be reproduced. 

Even the earliest semi-analytic models included feedback from supernovae in order to suppress the efficiency of galaxy formation and flatten the low-mass end of the luminosity function. The details of how winds are driven and stalled and thus the time-scale over which the gas ejected from the galaxy can return has varied. Some models assume the ejected ISM gas becomes part of the hot halo, where it immediately becomes available for cooling and re-accreting onto the galaxy, whereas others assume it gets ejected into an external IGM reservoir for a certain amount of time, before it rejoins the hot halo \citep[e.g.][]{Lucia2004}. There are also models that assume that the gas that escapes from the halo and reaches the Hubble flow never accretes onto the halo or galaxy again \citep{Bertone2007}.

Although these models do not directly trace wind recycling back into the galaxy, it can be informative to take a look at the time it takes before the ejected gas is reincorporated into the hot halo. \citet{Henriques2013} used Monte Carlo Markov Chain methods to find a redshift-independent reincorporation time-scale that is inversely proportional to halo mass: $t_\mathrm{reinc}=1.8\times10^{10}\,(10^{10}$~M$_\odot/M_\mathrm{halo})$~yr. This means that dwarf galaxies with $M_\mathrm{halo}\lesssim10^{10}$~M$_\odot$ retain few of their ejected baryons and therefore have little wind recycling, whereas massive galaxies with $M_\mathrm{halo}\gtrsim10^{12}$~M$_\odot$ have very short reincorporation times and the halo baryon fraction increases with halo mass, as observed \citep[e.g.][]{Lin2012}. Even though there is no explicit redshift dependence, ejection is most efficient at high redshift for halos at fixed mass, because cooling, star formation, and feedback are stronger on average. This results in a suppression of the formation of low-mass galaxies at early times and a more efficient build-up at later times, as observed via the galaxy stellar mass function. We will compare this time-scale to those derived from other models in Section~\ref{sec:concl}.

\subsection{Equilibrium models}

The interpretation of semi-analytic models and cosmological, hydrodynamical simulations can be difficult, because of their high level of complexity. An alternative approach is to employ simplified models based on the idea that galaxies self-regulate their growth, in a way that gas inflow and outflow approximately balance \citep[e.g.][]{Finlator2008, Bouche2010, Dave2012, Lilly2013, Dekel2014}. This is in part motivated by the observation that star-forming galaxies lie on a fairly tight relation between their SFR and stellar mass, which means that there appears to be an approximate equilibrium between gas supply and gas removal \citep[e.g.][]{Noeske2007, Lee2015}.

Equilibrium models start with the assumption that the equilibrium between accretion, star formation, and galactic winds evolves slowly. Galaxy mergers are generally not taken into account, but could set the scatter in galaxy scaling relations \citep{Mitra2017}. The accretion rates are based on those derived at the virial radius from dark matter-only simulations and modified by a mass-dependent parameter that quantifies the fraction of this gas that accretes onto a galaxy. Gas ejected from the ISM is usually parameterized as a mass loading factor in units of the galaxy's SFR. Recycling can be taken into account by either reducing the mass loading factor \citep[e.g.][]{Dekel2014} or adding an additional term to the accretion rate \citep[e.g.][]{Mitra2015}. The recycling material is assumed to have the same metallicity as the gas in the ISM. 

Matching the model parameters to reproduce stellar masses, SFRs, or metallicities in observations or simulations can give a handle on the importance of wind recycling. \citet{Finlator2008} find that roughly half of the accreting gas has gone through wind recycling at $z=2$. \citet{Dekel2014} find that even with their most efficient wind recycling, SFRs at $z=2$ are still under-predicted and argue for a more accurate treatment of the re-accretion of gas. \citet{Mitra2015} find that they can match observables well with parameters that depend on halo mass and redshift. They find that the gas recycling time-scale decreases with increasing halo mass (or stellar mass) and increasing redshift, but with a relatively weak dependence, $t_\mathrm{recycle}\propto M_\mathrm{halo}^{-0.45}$. We will discuss this time-scale further in Section~\ref{sec:concl}.

\subsection{Hydrodynamical simulations}

Full hydrodynamical, cosmological simulations broadly exist in two categories. One in which the volume is discretized (mesh-based simulations) and one in which the mass is discretized (particle-based simulations). Particle-based simulations have the ability to trace the trajectory of gas parcels and can therefore determine unambiguously when gas returns to a galaxy after having been ejected. Simulations show an increase in the gas fraction within $R_\mathrm{vir}$ towards more massive halos \citep[e.g.][]{Crain2010a, McCarthy2016, Voortetal2016}. This is a direct consequence of feedback removing gas more efficiently from low-mass halos. This already provides evidence for the increase of wind recycling in more massive galaxies. 

Cosmological simulations have limited resolution and therefore have to employ sub-grid models in order to simulate the large-scale effects of much smaller scale phenomena, such as supernovae or active galactic nuclei (AGN). The parameters of these models are not known a priori, but can be tuned in order to reproduce certain observables, such as the galaxy stellar mass function \citep[e.g.][]{Vogelsberger2014, Schaye2015}, although there may be many degeneracies. The importance of wind recycling depends strongly on the chosen sub-grid feedback model. Winds that are generated with a higher velocity, momentum, or temperature will travel farther away from the galaxy. Winds that are experiencing full hydrodynamical forces, such as ram pressure, will slow down more than those that have been `decoupled' and travel ballistically for some distance or time \citep[e.g.][]{Oppenheimer2010}. These differences likely result in the variety of recycling times found in the literature, as discussed in Section~\ref{sec:concl}.

Another interesting question arises when we consider gas that was near the center of the halo and would have accreted onto the galaxy, but was pushed out by galactic outflows. This gas can later accrete, but since it was never actually part of any galaxy, it is not considered wind recycling. However, one could argue that the difference between `true' wind recycling that originates from a galaxy and the swept-up gas around the galaxy that accretes at a later time is somewhat artificial. However, the gas will be more enriched if it is ejected from the ISM and therefore have a different impact on galaxy formation.

\section{The importance of feedback and wind recycling} \label{sec:results}

Using hydrodynamical simulations it is possible to follow a parcel of gas as it flows into and out of a galaxy. As mentioned above, some studies consider gas to be wind recycling when they accrete onto the same galaxy for the second time, whereas others consider this the case when they re-accrete onto any galaxy. Usually studies that use zoom-in simulations to focus on individual galaxies follow the former approach, whereas full cosmological simulations tend to study wind recycling statistically and therefore follow the latter approach.

\begin{figure}[!ht]
\center 
\includegraphics[scale=.5]{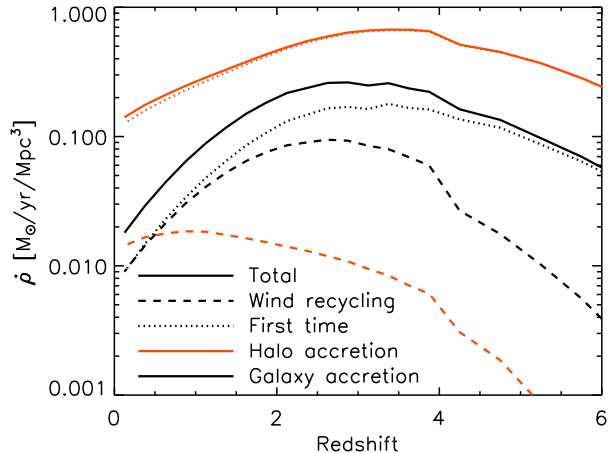}
\caption{\label{fig:global} Evolution of the global accretion rate densities in a (137~Mpc)$^3$ cosmological simulation with supernova and AGN feedback. The solid, black (red) curves show global accretion rate densities onto all galaxies (halos). Dashed curves show accretion rate densities resulting from wind recycling, dotted curved from gas that was never before in a galaxy's ISM (`first time'). The small `step' visible at $z\approx4$ is caused by the sudden increase in the time resolution for determining accretion (the time between snapshots decreases by a factor of 2 at $z=4$). The fraction of accretion onto halos due to wind recycling increases to almost 10 per cent at $z=0$ and is therefore clearly subdominant. At high redshift, wind recycling is also negligible for gas accretion onto galaxies. However, because of outflows driven by supernovae and AGN, its importance increases over time and rivals first time accretion at $z\approx0$. Adapted from \citet{Voort2011b}.}
\end{figure}
Figure~\ref{fig:global} shows the global accretion rate density onto halos (solid, red curve) and galaxies (solid, black curve), i.e.\ the gas mass accreting onto halos and galaxies, respectively, per year and per comoving Mpc$^3$. These global accretion rate densities are averaged over the time interval between two snapshots. The global accretion rate onto halos is substantially higher than the accretion rate onto galaxies, indicating that most of the gas that accretes onto halos never participates in star formation.

We split the global accretion rate into separate contributions from gas that was never in a galaxy before (here labeled `first time) and gas that was previously in a galaxy's ISM, but was ejected (`wind recycling'). Note that although the wind recycling gas is re-accreting, this could be onto a different galaxy than the one it was ejected from. Accretion onto halos is dominated by first time accreting gas at all times. Although accretion through wind recycling does become more important towards lower redshift, its accretion rate density is still 1~dex lower at $z=0$. This means that the direct effect of feedback and gas processing in galaxies are only of minor importance for accretion onto halos. At very high redshift, the first time accretion rate density onto galaxies is about an order of magnitude higher than the wind recycling accretion rate density. This difference decreases gradually and vanishes by $z=0$, even though first time accretion still dominates the growth of halos by 1~dex. This means that accretion onto galaxies through wind recycling happens efficiently and the majority returns to the galaxy without first leaving the halo. 

Once the gas is in the ISM, wind recycling and first time accreted gas are treated equally. The global SFR density therefore follows the galaxy accretion rate density, although with a delay resulting from the time it takes to convert the interstellar gas into stars. The gas can, however, be removed from the ISM by supernova and AGN feedback, as well as by dynamical processes. This is why the overall normalization of the SFR is generally lower than the ISM accretion rate.

\begin{figure}[!ht]
\center 
\includegraphics[scale=1.]{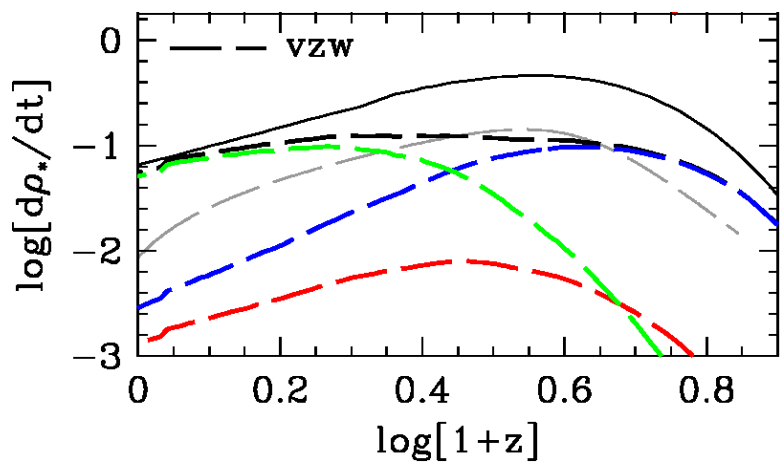}
\caption{\label{fig:globalstar} Evolution of the global SFR densities in a (69~Mpc)$^3$ cosmological simulation with (dashed, black curve) and without (solid, black curve) efficient stellar feedback (named `vzw'). The dashed gray curve shows the global SFR rate density derived from observations \citep{Hopkins2006}. Colored curves add up to the dashed, black curve. Green curves show the global SFR density resulting from gas that has been accreted multiple times (wind recycling). Blue (red) curves show the global SFR density from gas that has only accreted once with a maximum past temperature of less (more) than $2.5\times10^5$~K. At high redshift, most stars form from gas accreted for the first time, as also seen in Figure~\ref{fig:global}. The importance of wind recycling increases with decreasing redshift and it dominates by about an order of magnitude at $z=0$. This difference with Figure~\ref{fig:global} is due to a different treatment of the sub-grid feedback and the subsequent evolution of the ejected gas. Adapted from \citet{Oppenheimer2010}.}
\end{figure}

Figure~\ref{fig:globalstar} shows the evolution of the global SFR densities for a model without (solid, black curve) and with (dashed, black curve) efficient stellar feedback \citep{Oppenheimer2010}. Green curves show the global SFR density resulting from gas that has been accreted multiple times (wind recycling). Blue (red) curves show the global SFR density from gas that has only accreted once with a maximum past temperature of less (more) than $2.5\times10^5$~K. Wind recycling becomes more important at later times, in agreement with Figure~\ref{fig:global}. It dominates the global SFR after $z=1.7$ and is about an order of magnitude higher than first time accretion at $z=0$. This simulation leads to the conclusion that the vast majority of the late-time gas accretion onto galaxies (and resulting star formation) is gas that was previously ejected from a galaxy. This difference with Figure~\ref{fig:global} is due to a different treatment of the sub-grid feedback and the subsequent evolution of the ejected gas. Compared with the observed SFR density \citep{Hopkins2006} (gray dashed lines), the feedback model in Figure~\ref{fig:globalstar} over-predicts the amount of star formation after $z\approx1$, which is primarily a result of excessive star formation in massive galaxies. Suppressing some of the low redshift gas accretion and star formation with more efficient feedback could reduce the discrepancy with Figure~\ref{fig:global}, although it is unlikely to resolve it completely and shows how sensitive the quantitative results are to feedback implementations. 

One can also pose the reverse question, not what fraction of accreted gas accretes through wind recycling, but what fraction of gas ejected from galaxies and never re-accreted. \citet{Christensen2016} find that this is about 50 per cent of the mass in their zoom-in simulations (with $M_\mathrm{halo}<10^{12}$~M$_\odot$).

\subsection{Ejective and preventive feedback}

Naively one would expect feedback from stars and black holes to reduce the amount of gas in the ISM and potentially also reduce the gas accretion rate (through the interaction between inflowing and outflowing gas). However, this is not necessarily the case because as gas is ejected or prevented from accreting at high redshift, it is still available for feeding galaxies at low redshift. In the absence of galactic winds, most baryons remain in the galaxy they first accreted onto. This means that a lot of baryons are already `locked up' at high redshift in relatively low-mass galaxies. However, if this gas is ejected from the galaxies (or never accreted in the first place) it remains available to fuel massive galaxies at low redshift \citep[e.g.]{Voort2011a, Faucher2011}. 

\begin{figure}[!ht]
\center 
\includegraphics[scale=.55]{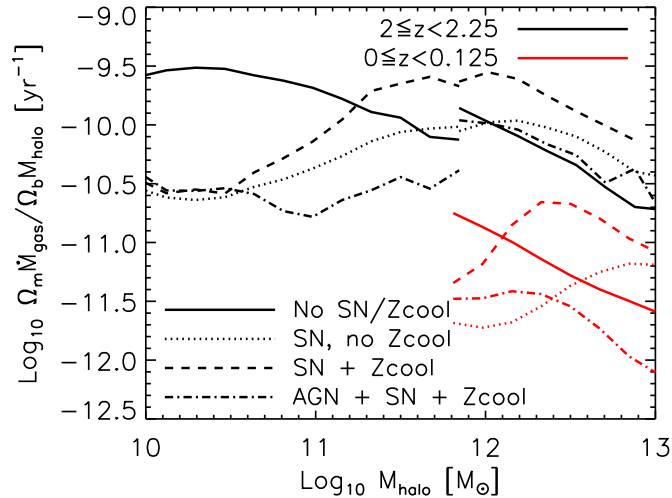}
\caption{\label{fig:accrate} Specific gas accretion rates onto galaxies against halo mass for different simulations at $z=2$ (black curves) and $z=0$ (red curves). The curves at low (high) halo masses are derived from simulations with higher (lower) resolution in smaller (larger) volumes. In the absence of supernova feedback (No SN/Zcool; solid curves), the specific accretion rate onto galaxies declines with halo mass, indicating that gas accretes less efficiently onto galaxies in higher mass halos. Efficient supernova feedback (SN, no Zcool; dotted curves) reduces the accretion rates by up to an order of magnitude for galaxies in low-mass halos, resulting in a peak in the specific gas accretion rate at $M_\mathrm{halo}\approx10^{12}$~M$_\odot$. It increases the accretion rates at high masses, because less gas has been locked up in smaller galaxies at higher redshift and is still available for accretion at low redshift. Including metal-line cooling (SN + Zcool; dotted curves) increases the accretion rate at all masses, but most strongly for $M_\mathrm{halo}\approx10^{12}$~M$_\odot$ ($T_\mathrm{vir}\approx10^6$~K). Efficient AGN feedback (AGN + SN + Zcool; dot-dashed curves) reduces the accretion rates for the highest halo masses, also by up to an order of magnitude. Adapted from \citet{Voort2011a}.}
\end{figure}

In Figure~\ref{fig:accrate} the specific accretion rate onto central galaxies as shown as a function of halo mass for 4 different simulations at $z=2$ (black curves) and $z=0$ (red curves). The curves at low (high) halo masses are derived from simulations with higher (lower) resolution in smaller (larger) volumes. The solid curves show simulations without supernova feedback and metal-line cooling, the dotted curves include supernova feedback, but no metal-line cooling, the dashed curves include both supernova feedback and metal-line cooling, and the dot-dashed curves use simulations that additionally include AGN feedback (as in Figure~\ref{fig:global}). 

The specific accretion rates onto galaxies decreases with halo mass in the absence of galactic winds (even though the specific accretion rates onto halos stays roughly constant), because cooling becomes less efficient at higher temperatures (in more massive halos). Supernova feedback reduces the galaxy accretion rate onto low-mass galaxies strongly by up to an order of magnitude. On the other hand, it actually increases the accretion rate onto massive galaxies, because less gas has been locked up in smaller galaxies at earlier times. The inclusion of metal-line cooling enhances the gas accretion rate, showing that, besides gravity, cooling sets the rate at which gas can accrete. AGN feedback reduces the galaxy accretion rate onto high-mass halos strongly, again by up to an order of magnitude. In these feedback models, the specific galaxy accretion rate peaks around $10^{12}$~M$_\odot$, where galaxy formation proceeds most efficiently. 

Internal feedback processes are not the only ones that can prevent gas accretion onto galaxies. External processes can do this as well. The environment of a galaxy is thought to be primarily determined by properties of the dark matter-dominated halo it lives in \citep[e.g.][]{Crain2009, Blanton2007, Wilman2010}. The removal of a satellite galaxy's gaseous halo can result in the suppression of gas accretion (and a resulting decline of the SFR), also known as `starvation' or `strangulation' \citep[e.g.][]{Larson1980, Balogh2000}. Using hydrodynamical simulations, \citet{Simha2009} indeed found lower gas accretion rates for satellites in massive halos compared to centrals of the same mass. The suppression of the gas accretion rate increases with host halo mass \citep{Voort2017}. ISM gas lost through ram pressure stripping or galactic winds will also be less likely to re-accrete onto satellites, but is an important fraction of the total gas accretion onto central galaxies at low redshift \citep{Angles2017}. 

\begin{figure*}
\begin{center}
\includegraphics[scale=0.45]{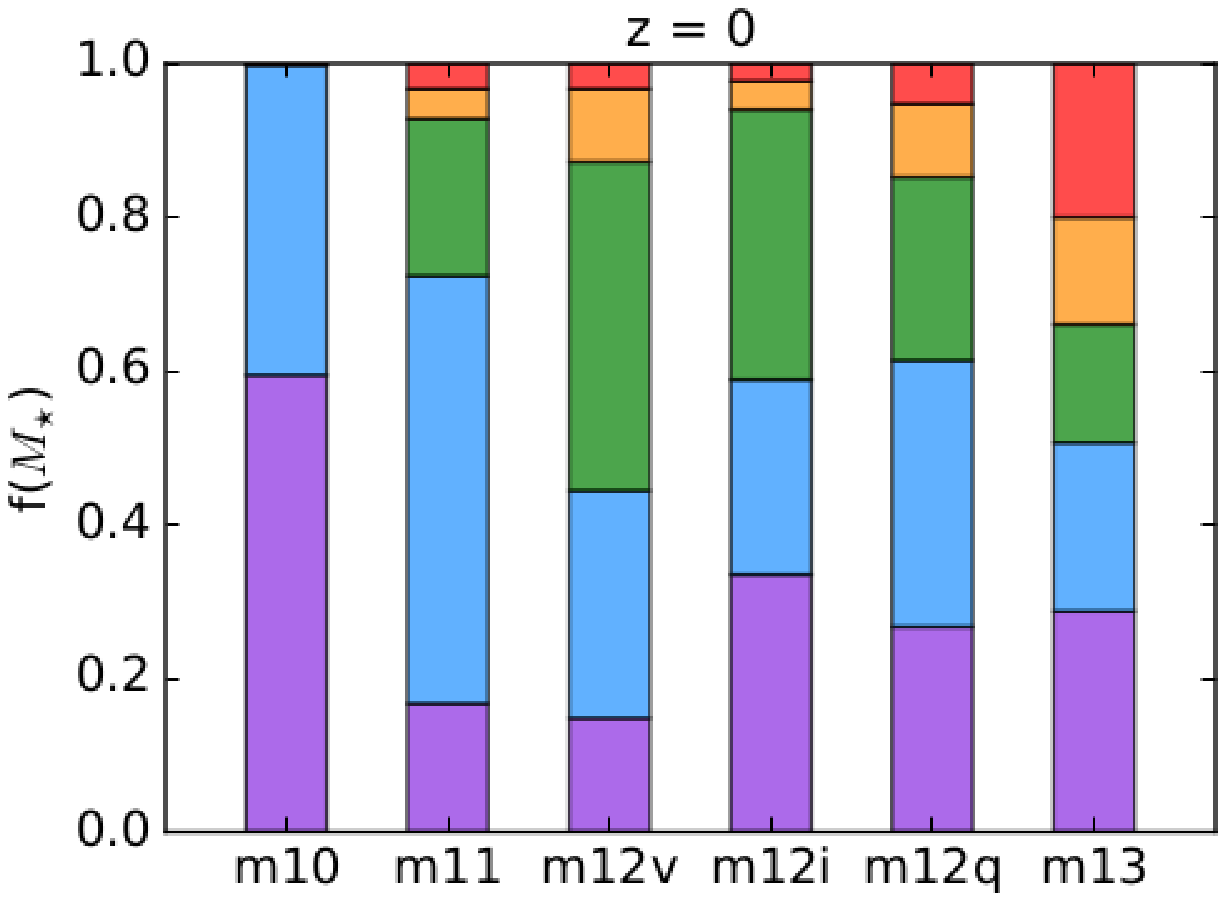}
\includegraphics[scale=0.45]{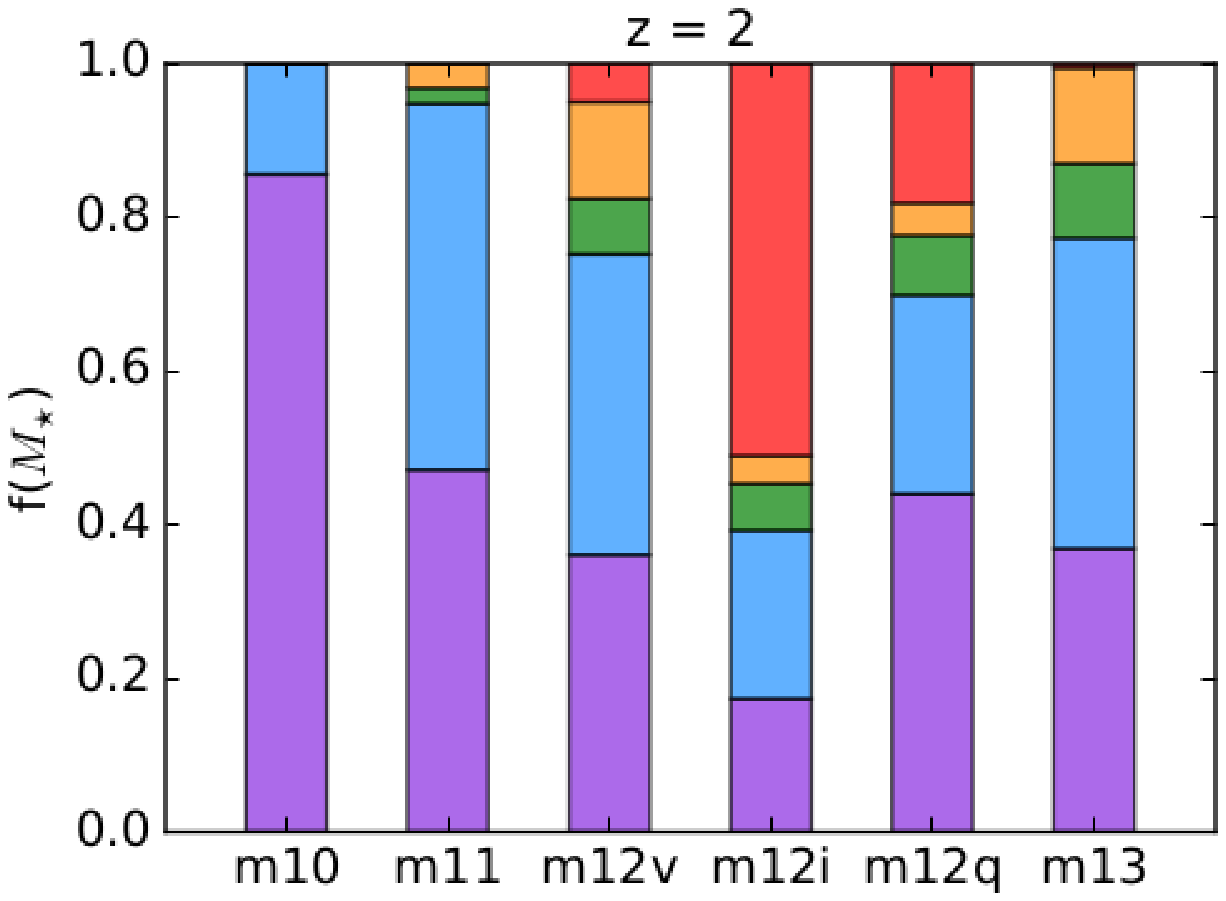}
\end{center}
\caption{\label{fig:AA17} Fraction of stellar mass at $z=0$ (left) and $z=2$ (right) contributed by gas accreted only once (purple), gas re-accreting onto the same galaxy (blue), gas re-accreted onto a different galaxy (green), gas accreted via galaxy mergers (orange), and stars accreted via galaxy mergers (red) for 6 simulated, central galaxies in order of increasing halo mass at $z=0$ (from $7.8\times10^{9}$ to $1.4\times10^{11}$ to $6.3\times10^{11}$ to $1.1\times10^{12}$ to $1.2\times10^{12}$ to $6.1\times 10^{12}$~M$_\odot$). Gas accreted for the first time is responsible for less than half the stellar mass for all galaxies except the lowest mass dwarf. The re-accretion of gas becomes more important towards lower redshift, especially gas processed and ejected in external galaxies (mostly satellites) before accreting onto the central and forming stars. Adapted from \citet{Angles2017}}
\end{figure*}

Figure~\ref{fig:AA17} shows the fraction of the stellar mass formed in different ways for individual central galaxies from cosmological zoom-in simulations at $z=0$ (left) and $z=2$ (right). From bottom to top: gas that accreted for the first time is shown in purple, gas that re-accreted onto the galaxy after being ejected from the same galaxy is shown in blue, gas that accreted onto the central after having been ejected from surrounding (satellite) galaxies is shown in green, gas that is brought in through galaxy mergers is shown in orange, and stars formed in other galaxies and accreted through mergers is shown in red. Gas previously ejected from external galaxies becomes more important with decreasing redshift, while the fraction of gas accreted only once decreases. The fraction of gas that accreted onto the same galaxy is an important path through which galaxies grow at both high and low redshift.

\subsection{The effect of feedback on the properties of accreting gas} 

Gas accretion is difficult to observe directly. Besides correctly reproduced galaxy masses and SFRs, one major test of galaxy formation models is whether or not they can reproduce the metallicity distribution in the Universe. This is one of the main reasons why wind recycling is of great importance for understanding galaxy evolution. The distance out to which ejected gas travels determines the metallicity of the CGM and IGM. The time-scale on which it re-accretes onto galaxies plays an important role in setting the metallicity of the ISM and of the resulting stars that form. For example, a decrease of the infall of pristine gas can explain the fact that the metallicity is observed to increase with decreasing SFR at fixed stellar mass \citep{Mannucci2010, Dave2011, Peng2015, Lagos2016, Kacprzak2016, Bahe2017}. 

Gas around galaxies has been detected in observations in absorption in quasar sightlines and in emission in hydrogen Lyman-$\alpha$, O\,\textsc{vi}, and soft X-rays \citep[e.g.][]{Peroux2005, Turner2016, Steidel2011, Hayes2016, Anderson2015}. However, it is difficult to determine whether this gas is accreting, outflowing, or static. Comparing results to hydrodynamical simulations is essential for their interpretation. \citet{Voort2012} used simulations to show that the vast majority of high-column density H\,\textsc{i} absorbers ($N_\mathrm{H\,\textsc{i}}\gtrsim10^{17}$~cm$^{-2}$) at $z=3$ are associated with galaxy halos. 60-70 per cent of this neutral gas accretes onto a galaxy by $z=2$, so within a Gyr. However, only roughly half of the accreting gas did so for the first time. Accreting gas will therefore not necessarily be metal-poor and it can be difficult to separate inflows and outflows based on their metallicity, except in the most extreme cases \citep{Hafen2016}. 

\citet{Ford2014} find that only 3 per cent of the low-redshift CGM and IGM is re-accreting onto galaxies after having been ejected from a galaxy, yet this recycling material contains 35 per cent of the metal mass. The rest of the metal mass is primarily contained in previously ejected gas that is not accreting onto galaxies. The former accounts for most of the low-ionization metal-line absorbers, such as Mg\,\textsc{ii}, while the latter accounts for the high-ionization absorbers, since it generally has a higher temperature. This suggests that accreting gas has likely already been detected via low-ionization absorption and wind recycling gas via low-ionization metal-line absorption \citep{Ford2014}. This identification can perhaps only be achieved statistically, because disentangling these from cool outflows is difficult. However, there are observations that show enhanced Mg\,\textsc{ii} absorption along the major and minor axis of galaxies, potentially showing evidence for biconical outflows along the minor axis and recycled wind accretion along the major axis \citep{Kacprzak2012}. 

\begin{figure}[!ht]
\center 
\includegraphics[scale=.45]{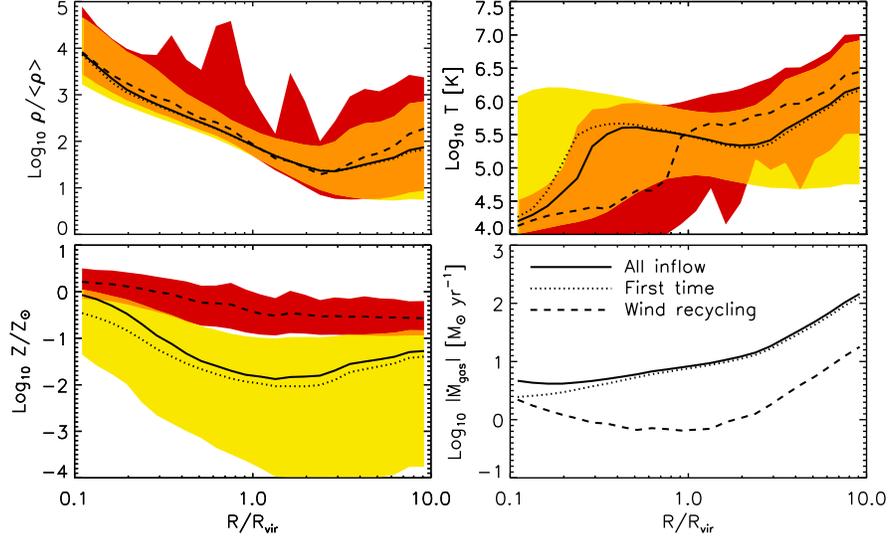}
\caption{\label{fig:radial} Properties of gas in halos with $10^{11.5}<M_\mathrm{halo}<10^{12.5}$~M$_\odot$ at $z=0$ as a function of radius for all inflowing gas (solid curves), gas flowing in for the first time (dotted curves) and wind recycling gas (dashed curves). The curves show the median values, except for the bottom, right panel which shows the mean value. Shaded regions show values within the 16th and 84th percentiles, i.e. the $\pm1\sigma$ scatter around the median, for first time accreting and wind recycling gas. From the top-left to the bottom-right, the different panels show the flux-weighted median gas overdensity, temperature, metallicity, and the mean accretion rate. A small fraction of the wind recycling gas has higher densities than average and most of it has a substantially lower temperature inside $R_\mathrm{vir}$. Wind recycling gas has much higher metallicity than gas accreting for the first time, especially around $R_\mathrm{vir}$. First time accreting gas dominates at all radii, but wind recycling contributes substantially in the centers of these halos and therefore to the accretion rate onto galaxies. Adapted from \citet{VoortSchaye2012}.}
\end{figure}

Figure~\ref{fig:radial} shows the flux-weighted median gas overdensity (top, left panel), temperature (top, right panel), metallicity (bottom, left panel), and mean accretion rate (bottom, right panel) of all inflowing gas, i.e.\ gas with negative radial velocity with respect to the central galaxy, as a function of radius for halos with $10^{11.5}<M_\mathrm{halo}< 10^{12.5}$~M$_\odot$ at $z=0$. The solid curves show the median values for all accreting gas, except for the bottom, right panel which shows the mean value. The dashed (dotted) curves show the properties for wind recycling (first time accreting) gas, i.e. gas which was (not) previously in the ISM of a galaxy. The shaded regions show values within the 16th and 84th percentiles. Wind recycling gas at radii larger than about $2R_\mathrm{vir}$ is dominated by gas associated with external galaxies and gas flowing towards our selected Milky Way-mass galaxies are likely flowing out of other galaxies.

At and beyond the virial radius, only about 10 per cent of accretion is due to wind recycling. This reflects the cosmic average fraction of gas ejected from galaxies by $z=0$. However, it becomes more important at smaller radii until it reaches about 50 per cent at $0.1R_\mathrm{vir}$, consistent with the global accretion rate density in Figure~\ref{fig:global}. The gas density is similar for accreting and re-accreting gas, although the latter has some high-density peaks of gas, which has likely recently been ejected from satellite galaxies. The differences are much larger for the gas temperatures. Inflowing wind recycling gas has a higher temperature outside $R_\mathrm{vir}$, likely because it has recently been heated by feedback from other galaxies. Inside $R_\mathrm{vir}$ the median temperature drops below the virial temperature and the difference between the temperature of gas flowing in for the first time and wind recycling gas is more than an order of magnitude around $0.4R_\mathrm{vir}$. Since the densities are similar, the temperature difference is primarily a result of the metallicity difference. At all radii, the metallicity of wind recycling gas is substantially higher than that of gas accreting for the first time. The difference is largest around $R_\mathrm{vir}$ ($\approx1.5$~dex), but is still large ($\approx0.7$~dex at $0.1R_\mathrm{vir}$. This shows that wind recycling is essential for accretion onto galaxies as well as for the temperature and metallicity structure of the CGM. 

The inclusion of galactic feedback can result in more realistic galaxies and enhance disc formation. Galactic outflows due to stellar winds originate from the sites of star formation. Since the SFR is highest in the centers of galaxies, ejected gas preferentially comes from the center as well and thus has low angular momentum. Disc formation can then proceed in one of two ways. Either low-angular momentum gas is removed from the galaxy completely, or the gas that re-accretes does so with higher angular momentum than when it was ejected. No consensus has been reached on which is more important. \citet{Ubler2014} find no change in angular momentum for re-accreting gas from the time it was ejected. In their simulations, feedback promotes disc formation by removing low-angular momentum gas at high redshift, allowing galaxies to be dominated by late-time gas accretion with high angular momentum. However, \citet{Brook2012} and \citet{Christensen2016} find that gas returns with significantly higher angular momentum than when it was ejected, because the ejected gas absorbs angular momentum from the CGM.

\section{Discussion and conclusions} \label{sec:concl}

This chapter reviews the effect of feedback on the accretion of gas, how it ejects material from the ISM and simultaneously prevents gas in the CGM (and IGM) from accreting onto the galaxy (and halo). Besides this negative effect on gas accretion, simulations show a positive effect as well, especially at late times and for massive galaxies. Gas that would have otherwise been locked up in galaxies becomes available for accretion at a later time, because of feedback. Furthermore, metal enrichment also enhances cooling, which further leads to enhanced gas accretion. We further focus on the accretion of gas that has previously been ejected from a galaxy and therefore has different properties than the gas that accretes directly from the IGM.

Mixing between hot outflows and the cooler CGM may be artificially suppressed in particle-based simulations and artificially enhanced in grid-based simulations or semi-analytic models. In reality, instabilities at the interfaces between the volume filling heated gas and embedded cool `clouds'  can mix the different gas phases \citep[e.g.][]{Veilleux2005}. Properly resolving these dynamics is challenging for simulations of isolated galaxies or idealized patches of galactic discs \citep[e.g.][]{Martizzi2016}, let alone for cosmological simulations. The inclusion of magnetic fields and/or thermal conduction may also be critical for capturing the correct dynamics at the interface between the cool and hot phases \citep[e.g.][]{McCourt2015, Bruggen2016}. 
In most particle-based simulations, metals are locked into gas particles. Either grid-based calculations or (particle-based simulations with explicit and realistic metal diffusion) would be useful for determining the extent to which small-scale mixing modifies the temperature, density, and, especially, metallicity of the ejected gas and thus its re-accretion. In the absence of such mixing the gas metallicity cannot converge at low metallicity. 
Thus, it is possible that the re-accretion of gas would change at much higher resolution and with the inclusion of additional physics.

\begin{figure}[!ht]
\center
\includegraphics[scale=.55]{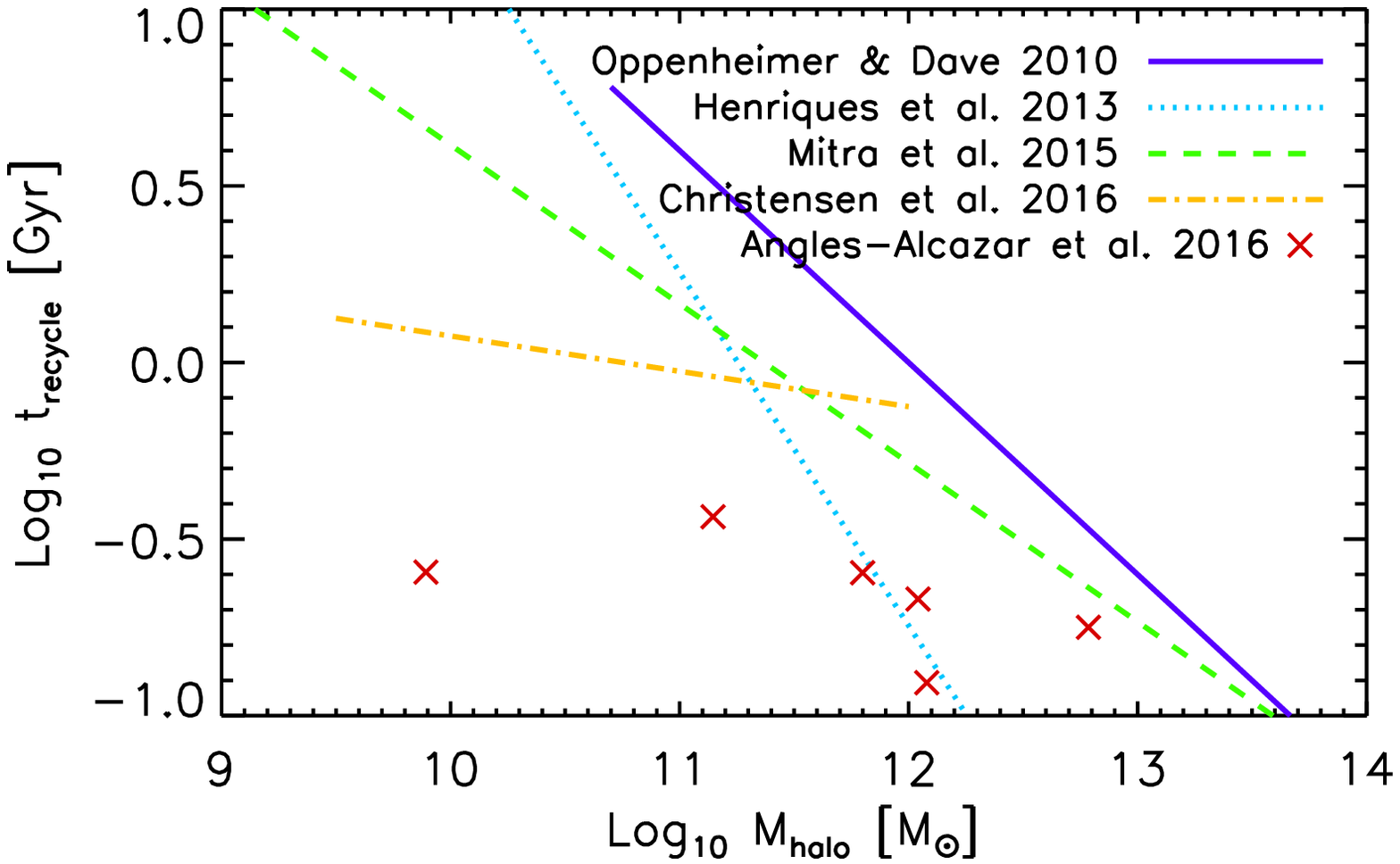}
\caption{\label{fig:trec} Literature compilation of recycling time-scales as a function of halo mass from \citet{Oppenheimer2010, Henriques2013, Mitra2015, Christensen2016}, and \citet{Angles2017}, as indicated by the legend. Most studies find relatively low recycling time-scales for massive galaxies, which means that ejected gas gets re-accreted fairly rapidly. Differences between studies are very large at the low-mass end, with some predicting that gas returns on a time-scale comparable to the age of the Universe, whereas others find time-scales similar to those at the high-mass end. Each of the studies shown employs a different definition for wind recycling (see also Section~\ref{sec:methods}). In reality, differences may therefore be somewhat smaller, but are unlikely to disappear.}
\end{figure}
Feedback is essential for bringing stellar masses into agreement with observations and for producing realistic galaxies. All methods discussed in Section~\ref{sec:methods} (semi-analytic models, equilibrium models, and hydrodynamical simulations) show that the re-accretion of gas after its ejection from a galaxy's ISM is an important ingredient for the growth of galaxies and for their metallicity evolution. However, there are large differences between estimates for its fractional contribution to galaxy growth and the time-scale for wind recycling. This can be understood by realizing that in order to reduce the ISM and stellar masses, one can either prevent more gas accretion onto the ISM or eject more gas from the ISM. Additionally, gas can be removed for a long time after a single ejection or be ejected many times if it re-accretes on a short time-scale.

Figure~\ref{fig:trec} shows a compilation of wind recycling time-scales from the literature \citet{Oppenheimer2010, Henriques2013, Mitra2015, Christensen2016, Angles2017}. Different studies indeed find different time-scales for the re-accretion of gas onto galaxies, which scale differently with galaxy or halo mass. At the high-mass end, there is more agreement that gas re-accretes on a fairly short time-scale. At the low-mass end, however there is a wide variety of predicted time-scales, some similar to those at the high-mass end and other similar to the age of the Universe. Each of the studies shown employs a different definition for wind recycling and differences may therefore be somewhat smaller in reality. However, even with the same definition, these discrepancies will likely remain, because of the very different feedback implementations. 

Although these different feedback implementations may result in similar galaxy masses, other galaxy and CGM properties can be very different. For example, gas cycles on short time-scales can affect the gravitational potential. The collisionless dark matter and stars in the halo center/galaxy respond to this changing potential, which results in dark matter cores and an increase in the effective radius of dwarf galaxies \citep{Pontzen2012, Badry2016}. \citet{Suarez2016} show in 2D grid simulations that the amount of cool gas formed in an outflow depends strongly on the feedback cycle as well. Shorter, stronger bursts produced by a rapidly fluctuating SFR generally lead to a larger fraction of cool gas forming in the outflow, although long feedback cycles (of a Gyr or more) also allow cool gas to form. This also clearly affects the structure of the CGM and the importance of re-accretion. Alternative feedback processes, not yet included in studies of wind recycling, such as cosmic ray feedback \citep[e.g.][]{Uhlig2012, Booth2013, Salem2014} could be important in driving outflows. For instance, cosmic rays may accelerate gas more smoothly than in the simulations described here, where feedback is more violent, and drive galactic winds with qualitatively different properties. 

This chapter aimed to show how outflows driven by galactic feedback change the rates and properties of gas accretion onto galaxies. Although there are still large uncertainties, it is clear that wind recycling is an important component of the growth of galaxies, especially at late times. If we wish to understand galaxy evolution, we need to understand how outflows and inflows interact.


\end{document}